\documentclass[aps,prb,superscriptaddress, showkeys, twocolumn]{revtex4-1}
\usepackage{amsmath,amssymb}
\usepackage{graphicx}
\usepackage{dcolumn}
\usepackage{bm}
\usepackage[latin1]{inputenc}
\usepackage{float}
\usepackage{color}
\usepackage{booktabs}
\usepackage{amsmath}
\definecolor{rojo}{rgb}{1,0,0}
\definecolor{verde}{rgb}{0,0.8,0.5}
\definecolor{azul}{rgb}{0,0,1}
\definecolor{rosa}{cmyk}{0,1,0,0}
\usepackage[toc,page]{appendix} 
\usepackage{latexsym}
\usepackage{amsmath}
\usepackage{mathrsfs}
\usepackage{bm}
\usepackage{amsthm}
\usepackage{physics}
\usepackage{adjustbox}
\usepackage{bbm}
\usepackage{amsfonts}
\usepackage{amssymb}
\usepackage{color}
\usepackage{epsfig}
\usepackage{hyperref}
\hypersetup{colorlinks=true, citecolor=blue}
\usepackage{soul}
\usepackage[normalem]{ulem}
\usepackage[cmtip,all]{xy} 
\usepackage{comment}
\usepackage{multirow}
\usepackage{mathtools}
\usepackage{epstopdf}
\newcolumntype{L}{>{$}l<{$}}

\newcommand{\longsquiggly}{\xymatrix{{}\ar@{~>}[r]&{}}} 

\begin{document}

\title{{\color{blue}  Radiation Modulated Spin coupling in DNA}}

\author{Alexander L\'opez}
\affiliation{Departamento de F\'isica, Escuela Superior Polit\'ecnica del Litoral, ESPOL, Campus Gustavo Galindo Km. 30.5 V\'ia Perimetral, PO Box 09-01-5863, Guayaquil, Ecuador}

\author{Solmar Varela}
\affiliation{Theoretical  Condensed Matter Group, School of Chemical Sciences \& Engineering, Yachay Tech University, 100119-Urcuqu\'i, Ecuador}

\author{Ernesto Medina}
\email{ernestomed@gmail.com}
\affiliation{Departamento de F\'isica, Colegio de Ciencias e Ingenier\'ia, Universidad San Francisco de Quito, Diego de Robles y V\'ia Interoce\'anica, Quito, 170901, Ecuador}


\date{\today}

\begin{abstract}
The spin activity in macromolecules such as DNA and oligopeptides, in the context of the Chiral Induced Spin Selectivity (CISS) has been proposed to be due to the atomic Spin-Orbit Coupling (SOC) and the associated chiral symmetry of the structures. This coupling, associated with carbon, nitrogen and oxygen atoms in biological molecules, albeit small (meV), can be enhanced by the geometry, and strong local polarization effects such as hydrogen bonding (HB). A novel way to manipulate the spin degree of freedom is by modifying the spectrum using a coupling to the appropriate electromagnetic radiation field. Here we use the Floquet formalism in order to show how the half filled band Hamiltonian for DNA, can be modulated by the radiation to produce a up to a tenfold increase of the effective SOC once the intrinsic coupling is present. On the other hand, the chiral model, once incorporating the orbital angular momentum of electron motion on the helix, opens a gap for different helicity states (helicity splitting) that chooses spin polarization according to transport direction and chirality, without breaking time reversal symmetry. The observed effects are feasible in physically reasonable parameter ranges for the radiation field amplitude and frequency.
\end{abstract}
\maketitle

\section{Introduction}
A number of spin related phenomena in amino-acids, oligopeptides, DNA and proteins have been the focus of recent interest\cite{NaamanAminoacids,Aragones,NaamanDNA,Gohler,NaamanPhotosystem}. Most of the related works have dealt with the high degree of spin polarization that can be achieved when injecting spin unpolarized electrons through chiral molecules\cite{Gohler,NaamanAminoacids,Aragones}. These observations are known as the Chiral Induced Spin Selectivity (CISS) effect, since chirality seems to be a crucial ingredient for spin filtering to occur. A variety of different mechanisms have been proposed that seem to explain qualitatively the effect\cite{ACSNanoReview,AharonyReview}. Recent works\cite{Varela2016,Varela2018,HeliceneMujica,OligoPeptides} inspired by models on graphene corrugation, nanotubes and fullerenes\cite{Huertas2006,Ando} have assessed the role of the geometrical arrangement, of electrons-bearing orbitals, on the Spin-Orbit Coupling (SOC). A striking role is demonstrated for the chiral geometry, property without which spin filtering does not occur\cite{MedinaLopez,Sina,Varela2016,MedinaGonzalez2015,Carmen}. The crucial role of geometry on SOC has also been demonstrated by structural deformations \cite{NaamanDNA,Varela2018,VarelaJCPHydrogen,OligoPeptides} where the SOC can be modulated by either stretching or compressing the chiral molecule. Details of the orbitals contributing to SOC can be resolved by deformation details\cite{Varela2018,OligoPeptides}.

The mechanism by which geometry and deformations modulate SOC is by changing the orbital overlaps that connect electron bearing overlaps via spin-active channels\cite{Varela2016}. Such modulation could also be achieved by the coupling to radiation of the appropriate intensity and frequency. In this work we explore the effects on the spectrum of a DNA like molecule, of a periodic excitation by an electric field (described as classical light). We build on the models for DNA presented in reference [\onlinecite{Varela2016}] in order to analyze the role of high frequency modulation of the associated spin filtering properties of the DNA helix. Modulation of spin activity is assessed by using the Floquet formalism which deals with the role of periodic driving. Our main motivation is to determine the interplay of circularly polarized radiation, and the spin activity induced by SOC. 

The radiation coupling model relies on the already experimentally accessed capabilities of periodic interactions to generate modulations on the electronic properties\cite{Foa1} of low dimensional systems such as graphene. The Floquet formalism has received a great deal of attention\cite{McIver}. Among the phenomena of periodically driven interactions, the control of gaps in graphene\cite{Foa1} was an exciting development with technological implications. Later, the same formalism led to predict laser induced chiral edge states in nanoribbons\cite{Foa2}.  We also highlight the generation of topological phases\cite{NP2011Lindner,Wang453} of which CISS may show some manifestations since spin-flip scattering is weakly suppressed by a SO gap between different helicity states\cite{Varela2016,MedinaGonzalez2015,NaamanWaldeck}.
These periodically driven topological effects, involving the angular momentum of circularly polarized radiation, could be contrasted to the results in a recent work involving the connection between DNA like chiral molecules and topological effects going beyond chiral-induced spin selectivity\cite{OAM2021}. In this work, the authors predict the so called orbital polarization effect which could induce spin-selective phenomena in inversion-breaking materials, even in the absence of chiral symmetry\cite{OAM2021}. 

Many physically relevant properties of the out of equilibrium system can be captured within the high frequency or so called off resonant regime, where the driving fields induce energy band renormalization in the quantum optics. Accordingly, we stress that our work will focus on frequency regimes larger than the kinetic energy contributions, which is the dominant energy scale in the static scenario. 

We find that, in the high frequency regime for circularly polarized light in the molecular axis, the half filled Hamiltonian for DNA exhibits a modulation of its kinetic term. When the SOC is introduced the spectrum exhibits a gap at zero radiation intensity that can be substantially enhanced by the parameter $\xi=eER/\hbar\Omega$, where $E$ is the electric field intensity and $\Omega$ its frequency, while $R$ is the nearest neighbor distance. This effect has the double role of effectively increasing the apparent SOC but also enhancing the helicity splitting between Kramers doublets, that will play and important role in spin selection in tunneling phenomena\cite{VarelaIskra}.

The paper is summarized as follows: In section II we review the double helix model of DNA\cite{Varela2016}. We then apply circularly polarized radiation along the axis of the double stranded molecule and find the effective Floquet Hamiltonian in the high frequency limit. In section III we specialize the Hamiltonian to half filling, as electron bearing orbitals are the $\pi$ electrons of the bases, and completely solve for the eigenvalues and eigenfunctions of the model. In section IV we show the helicity splitting effect and how circularly polarized radiation can enhance the gap tenfold within reasonable parameter values. We close with a summary and conclusions.


\section{model}
\begin{figure}
    \centering
    \includegraphics[height=9cm]{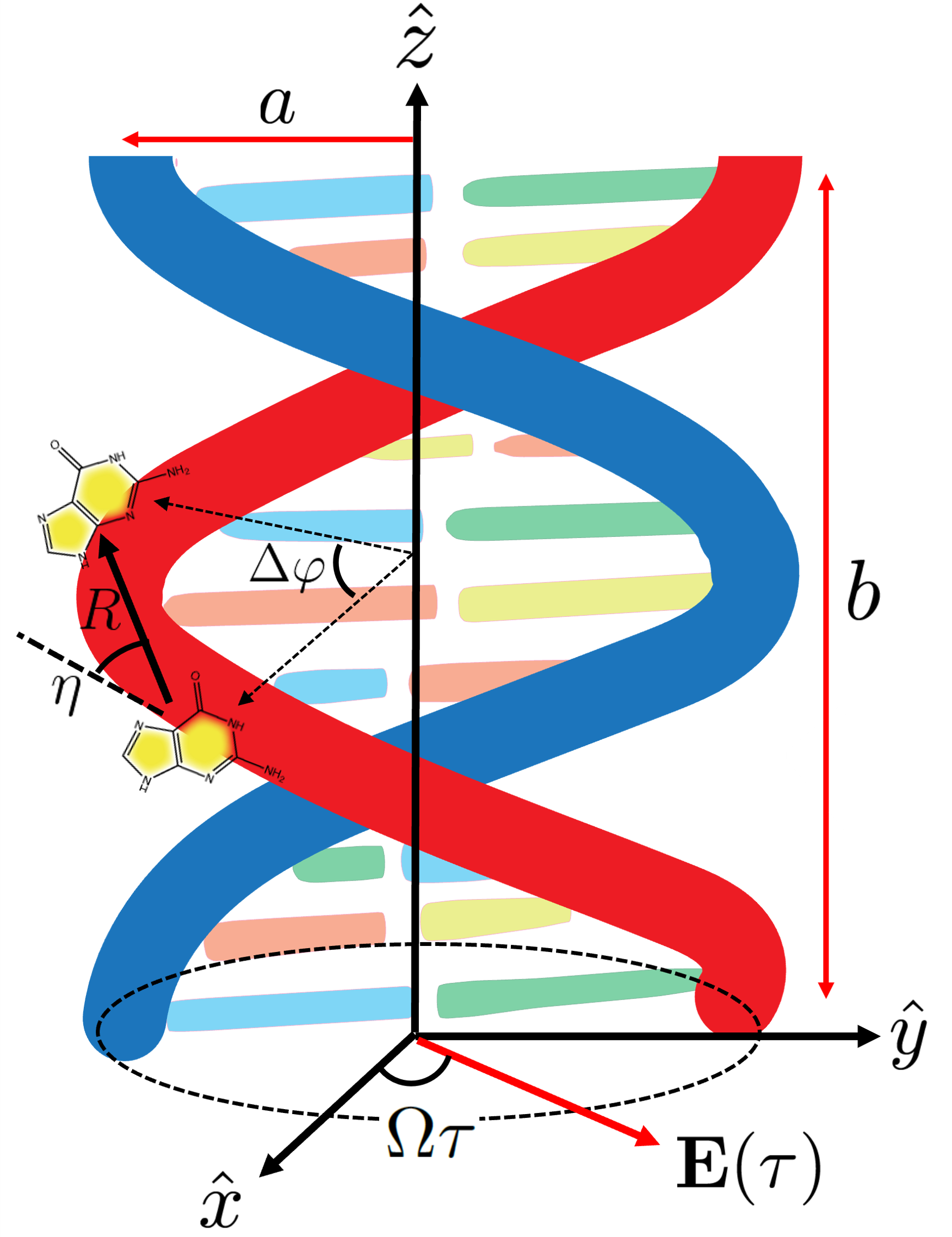}
    \caption{Schematic representation of the irradiated DNA molecule whose radius is $a$ and pitch $b$. Two consecutive bases in DNA are connected by the vector $\mathbf{R}=R(-\cos\eta\sin\varphi\hat {x}+\cos\eta\cos\varphi\hat y+\sin\eta\hat z)$. The radiation field $E(t)$ is in the xy plane, perpendicular to the helix z-axis, rotates at the rate $\Omega\tau$.}
    \label{Figure1}
\end{figure}
To model the system, we start from the effective static Hamiltonian for DNA in the absence of radiation within the tight binding formulation\cite{Varela2016}  
\begin{equation}
H_0=2t_if(k)\mathbbm{1}_\sigma\mathbbm{1}_s-2\lambda h(k,\tau)\mathbbm{1}_\sigma s_\varphi+t_o\sigma_r\mathbbm{1}_s, 
\end{equation}
where $f(k)=\cos\left(\mathbf{k}\cdot\mathbf{R}\right)$ and 
$h(k)=\sin(\mathbf{k}\cdot\mathbf{R})$, ~~$t_i$ ($t_o$) represents the intra (inter) strand first neighbor hopping term for the two helices that make up the DNA molecule, and $\lambda$ describes the effective intrinsic Spin-Orbit (SO) coupling strength. The vectors for the momentum operator and spin degree of freedom are represented by $\mathbf{k}$ and $\mathbf{s}$, respectively. The strand degree of freedom is represented as a pseudospin described by the Pauli matrices $\sigma_i$. In addition, $\mathbf{R}$ represents a vector connecting nearest neighbors on each DNA strand. From symmetry considerations, it is convenient to express the momentum vector in cylindrical coordinates, such that $\mathbf{k}=k_\varphi\hat{\varphi}+k_z\hat{z}$, with $\hat{\varphi}$ giving the unit tangent of the projected circle in the $xy$ plane and $\hat{z}$ along the helix axis.

\textcolor{black}{Upon inclusion of the radiation field, the Hamiltonian becomes time-dependent $H(t)=H_0+V(\tau)$, where the periodic driving interaction term $V(\tau)$ is periodic in the time parameter $\tau$, $V(\tau+T)=V(\tau)$.} 
\textcolor{black}{If the radiation impinges along the molecular axis, the associated electric field, will be perpendicular to the helix axis, and it is described as}
\begin{equation}
\mathbf{E}(\tau)=E[\cos(\Omega \tau)\mathbf{\hat{x}}+\sin(\Omega \tau)\mathbf{\hat{y}}],
\end{equation}
where $E$ and $\Omega$ are the amplitude and frequency of the radiation field, respectively. Positive (negative) frequency values correspond to right (left) circular polarization. Then, using the relation $\mathbf{E}=-\partial_\tau \mathbf{A}(\tau)$, the vector potential is found to be $\mathbf{A}(\tau)=E/\Omega(-\sin\Omega\tau\hat{x} +\cos\Omega\tau\hat{y})$. 
Using the minimal coupling prescription (i.e. Peierls substitution) $\mathbf{k}\rightarrow \mathbf{k}+e\mathbf{A}(\tau)/\hbar$, where $-e$ is the electron charge, and given the energy hierarchy for which the kinetic contribution is the largest relevant energy scale in the model, we get the effective time-dependent minimal Hamiltonian
\begin{equation}
H(k,\tau)=[2t_if(k,\tau)\mathbbm{1}_s\mathbbm{1}_s\sigma_x-2h(k)\lambda s_\varphi]\mathbbm{1}_\sigma+
t_o\mathbbm{1}_s\sigma_r
\end{equation}
where $\mathbbm{1}_{s(\sigma)}$ represents the $2\times 2$ identity matrix in the spin (pseudospin/strand) degree of freedom.  Moreover, we are neglecting the onsite energies and the direct modulation effects of the radiation field in the spin-orbit contribution. We have also introduced the tight-binding functions\cite{Varela2016}
\begin{eqnarray}
f(k,\tau)&=&\cos\left (\mathbf{k}\cdot\mathbf{R}-\xi\cos\eta\cos(\Omega \tau-\varphi)\right),\nonumber\\
\end{eqnarray}
in which $\mathbf{k}\cdot\mathbf{R}=R\cos{\eta}(k_{\varphi} + k_z\tan{\eta})$, being $R$ the lattice parameter and $\tan\eta=b/2\pi a$. We also have defined the effective light-matter contribution, via the dimensionless quantity $\xi=(eER/\hbar\Omega)$.

In the time-dependent Hamiltonian, the intra-strand hopping contribution to the kinetic energy can be expanded as
\begin{equation}
f(k,\tau)=\sum^{\infty}_{n=-\infty} f_n(k)e^{-in\Omega \tau},
\label{FourierSeries}
\end{equation}
such that
\begin{equation}
f_n(k)=\frac{1}{T}\int^T_0f(k,\tau) e^{in\Omega \tau}d\tau.
\end{equation}
Using the form for $f(k,\tau)$ and defining $\tau'=\Omega\tau$ we arrive at
\begin{equation}
f_n(k)=\frac{1}{2\pi}\int^{2\pi}_0\cos[\mathbf{k}\cdot\mathbf{R}+\xi\cos\eta\cos(\tau'-\varphi)] e^{in\tau'}d\tau'.
\label{fourierTransform}
\end{equation}
Then, the contribution from the lowest mode $n=0$, the dominant term in Eq.\ref{FourierSeries}, is $f_0(k)=\cos(\mathbf{k}\cdot\mathbf{R})J_0(\xi)$, and it is the leading contribution for the high-frequency regime $\hbar\Omega>t_i$, so that the Floquet side bands are sufficiently spaced and couple only weakly. 
The effective Hamiltonian is then 
\begin{equation}
\mathcal{H}=[2t_if_0(k)\mathbbm{1}_s-2\lambda h(k) s_\varphi]\mathbbm{1}_\sigma+
t_o\mathbbm{1}_s\sigma_r, 
\label{Heffec}
\end{equation}
where we emphasize the meaning of the three term: the first is the radiation modulated kinetic energy, the second is the SO term and the third is the inter-strand coupling which is spin inactive in the absence of a Rashba type term that is negligibly small at the field strengths we consider here due to radiation. The spin activity of the inter-strand term will only be appreciable due to internal electric fields of atomic origin\cite{Varela2016} or hydrogen bonding polarization\cite{VarelaJCPHydrogen}.

It is important to note the regimes of validity of this Hamiltonian in terms of the relative values of $t_i,\lambda$, the magnitude of the kinetic energy, the strength of the SO term and $\xi$ and the ratio of the electric field strength to its frequency $\Omega$. The regime in which we can stay with the lowest order term in the sum of modes (Eq.\ref{FourierSeries}) is that $\hbar\Omega > t_i$ while the $\xi$ is unrelated to this restriction and can take values where $J_0(\xi)$ undergoes changes in sign thus effective forcing changes in the direction of propagation of the electron. The values of $\lambda$ will always be two to three orders of magnitude smaller than the kinetic term, but are the only source of spin activity. Finally the inter-strand term of magnitude $t_o$ is spin inactive and much smaller than the kinetic term so that effectively transport occurs on the strands, uncoupled.

\begin{figure}[h]
  	  \centering
  	  \includegraphics[width=8.7cm]{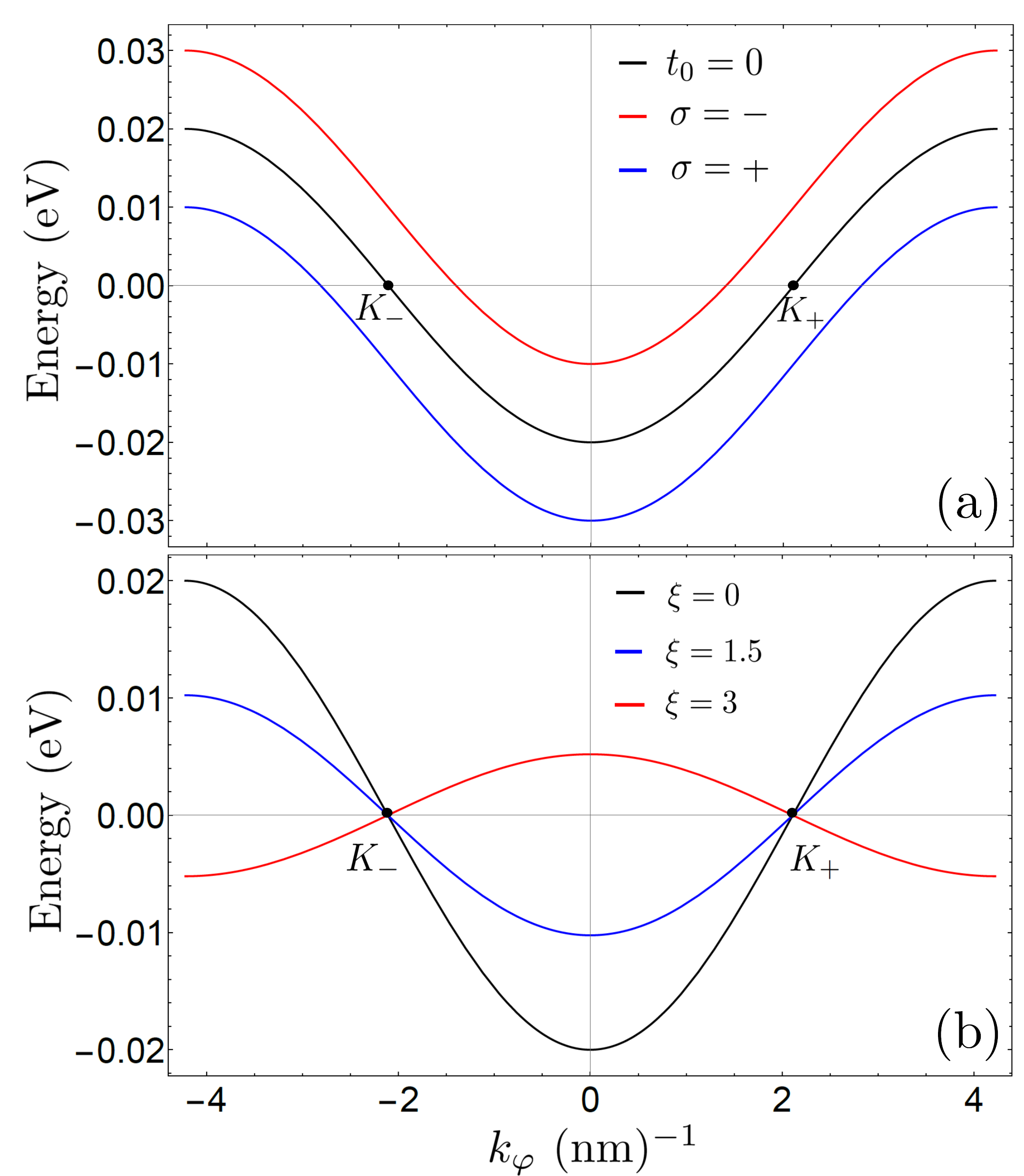}
  	  \caption{\color{black} Energy band versus the wave vector $k_{\varphi}$ in the first Brillouin zone. The top panel depicts inter strand effects, without radiation effects. The Bottom panel depicts radiation effect from three different $\xi=eER/\hbar\Omega$ values. Note that the velocities at the  symmetry points $K_{\nu}$ can be inverted by radiation effects. The half filling point is chosen to be at zero energy.}
  	  \label{EnergyVSk_2}
\end{figure}

\begin{figure}[h]
  	  \centering
  	  \includegraphics[width=8.7cm]{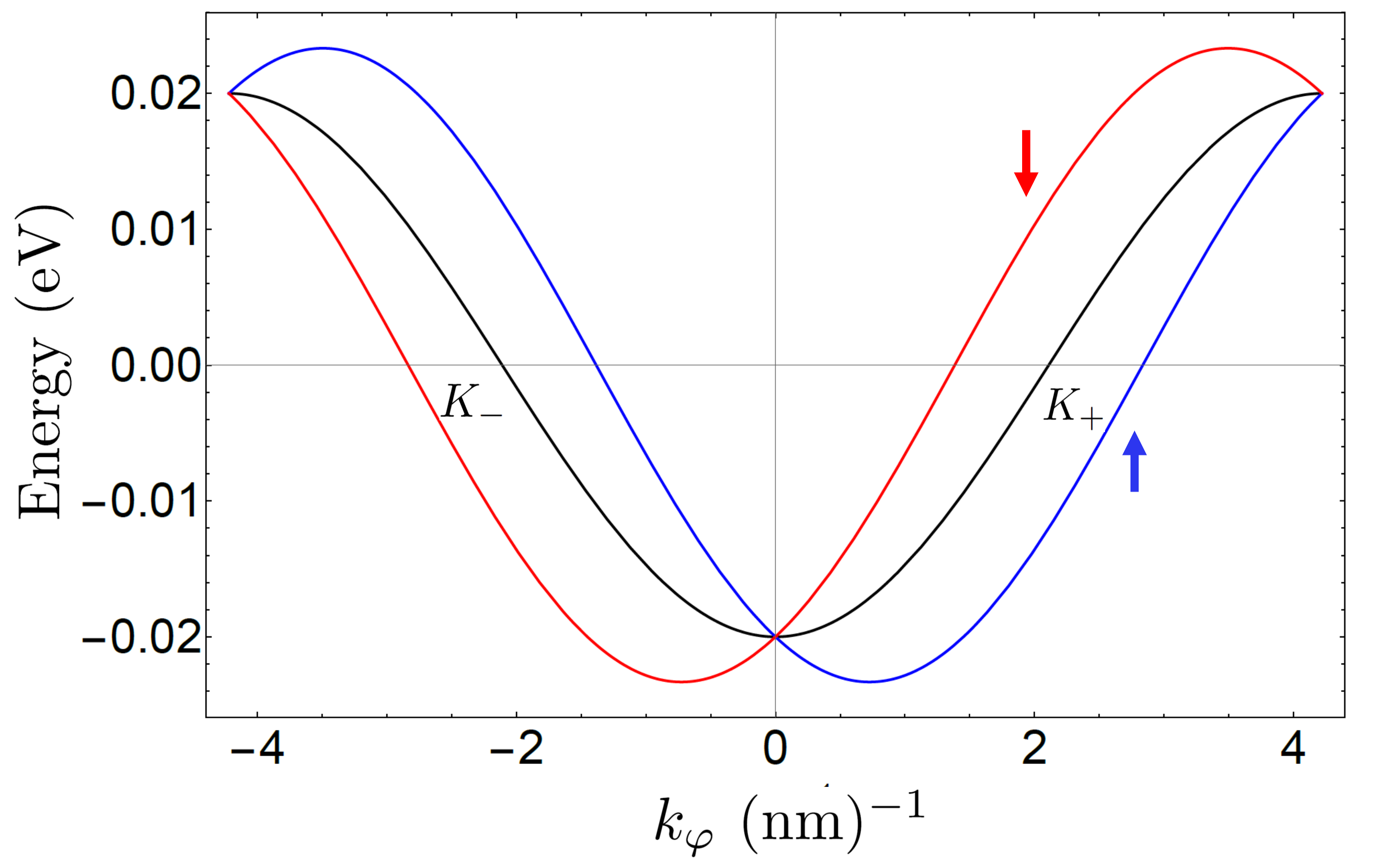}
  	  \caption{Energy band versus the wave vector $k_{\varphi}$ for $\nu=1$, when the spin-orbit coupling is on. The red (blue) lines correspond to the two helicity states parallel or antiparallel to the the $z$ axis.}
    \label{Figure3}
\end{figure}

Figure \ref{EnergyVSk_2} shows the simple band model for the $\pi$ coupled transport on DNA in the absence of the SO term. The top panel depicts the effects of the inter-strand term corresponding to the kinetic term as a function of $k_{\varphi}$. The $K_{\pm}$ denote the half filling of the band which is chosen to correspond to zero energy. The bottom panel depicts the effects of the field strength to frequency ratio $\xi$. A feature which will become critical in the later analysis is the possibility of inverting the electron velocity by the optical driving. Figure \ref{Figure3} depicts the spin split bands due to the SO term. In the next section we will see the behavior of the bands in the vicinity of half filling.

The ranges for the ratio $\xi=eER/\hbar\Omega$ considered in Fig.\ref{EnergyVSk_2} can physically be achieved for laser intensities of up to $130~mW/\mu m^2$ (see [\onlinecite{Foa1}]) yielding electric fields of $\sim 10^7 V/m$, with the values of $R\sim 4-9$~\AA, $t_i\sim 10~meV$\cite{Simserides}, $\hbar\Omega\sim 2~meV$ (infrared range) and $\lambda=6-20$ meV\cite{Huertas2006,Varela2016}.

\section{Half filling Hamiltonian}

Now, we focus on the half-filling continuum model, as transport electrons are in the nominally half filled $\pi$ orbitals of the bases\cite{Simserides}. 
Defining a small perturbation $\mathbf{q}$ around the half filling vectors $K_{\nu} = \nu\left(0,\frac{\pi}{2R\sec{\eta}},\frac{\pi}{2R\csc{\eta}}\right)$ with $\nu=\pm 1$, such that $\mathbf{k}=\mathbf{K_{\nu}}+\mathbf{q}$, we expand the Hamiltonian as 
\begin{eqnarray}\label{minimal}
&&\mathcal{H}=\\
&&-2\nu t_i J_{0}(\xi)(q_\varphi + q_z\tan{\eta})R\cos{\eta}\mathbbm{1}_s - 2\nu\lambda s_\varphi+t_o\sigma_r.\nonumber
\end{eqnarray}

\begin{figure}[h]
  	  \centering
  	  \includegraphics[width=8.7cm]{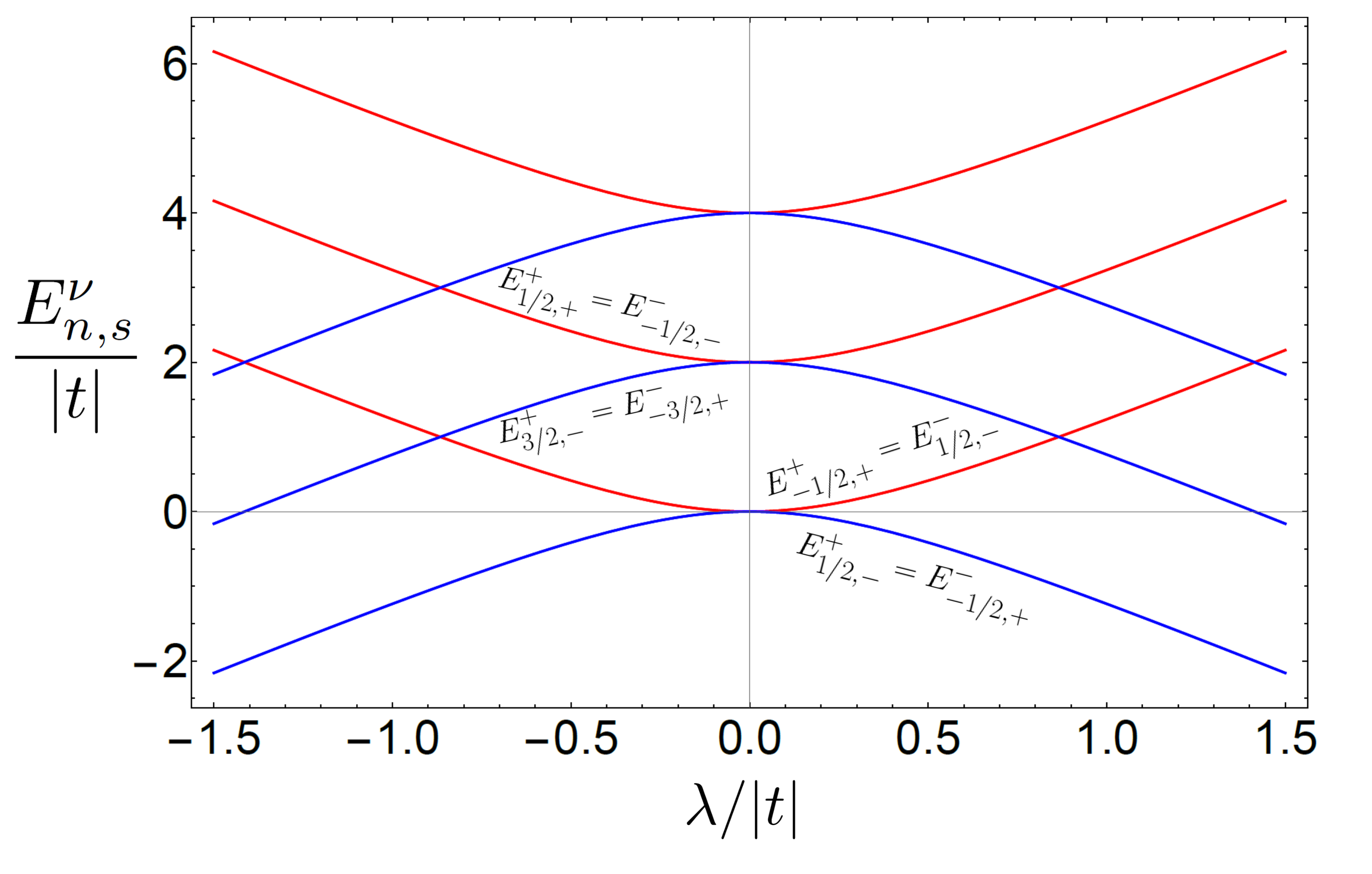}
  	  \caption{Energy of electrons as a function of SO interaction magnitude, $\lambda$ (in units of $t$), without effect of radiation.}
    \label{Figure4}
\end{figure}

Identifying $q_\varphi=-i\partial_\varphi/a$ ($a$ the radius of the double helix, see Fig.\ref{Figure1}) and \begin{equation}\label{mphi}
s_\varphi=\left(
\begin{array}{cc}
     0  &-ie^{-i\varphi}\\
ie^{i\varphi}     &0 
\end{array}
\right),
\end{equation}
and neglecting the inter-strand contribution, we obtain the minimal Hamiltonian that captures the modulation effects on the spin polarization in the vicinity of the $K$ points. Note that $[-i\partial_{\varphi},s_\varphi]\neq0$, but $[-i\partial_\varphi,J_z]=[s_\varphi,J_z]=0$, with $J_z=-i\hbar\partial_\varphi+\hbar s_z/2$, i.e. $J_z$ is the total angular momentum operator, comprising both the orbital and spin-dependent contributions. We thus find that $[J_z,H]=0$. Hence, we can rewrite the Hamiltonian (\ref{minimal}) as  \begin{equation}\label{minimal1}
H=- 2\nu (tJ_z+t_iRq_z\sin\eta)J_0(\xi)\mathbbm{1}_s+\nu(t s_zJ_0(\xi)-2\lambda s_\varphi). 
\end{equation}
with $t=t_i (R/a)\cos\eta$.
Choosing the  eigenbasis for angular momentum,
 \begin{equation}
J_z\ket{ns}=n\hbar\ket{ns},     
 \end{equation}
it follows that the spectrum of the minimal Hamiltonian is given
\begin{equation}
E_{n,s}^{\nu}=-2\nu (nt+t_iRq_z\sin\eta)J_0(\xi)+s\nu\sqrt{(tJ_0(\xi))^2+4\lambda^2},
\label{EnergyFull}
\end{equation}
with $s=\pm 1$ a spin index, $n \in \mathbb{Z}+1/2$, and $s\nu$ represents the helicity of the electron. {\color{black}Table \ref{table:1} shows the energies for the first three levels in the limit of $\lambda=0$ and how degeneracies are broken when the SO interaction is turned on. The spectrum is depicted in Fig.\ref{Figure4}. Note how, in the absence of SO coupling, the spectrum is fourfold degenerate, as required, as space and spin inversion symmetries are independent\cite{BercheChatelainMedina}}. On turning on the SO interaction, we only require time reversal symmetry and propagation direction and spin become coupled. The SO coupling breaks the fourfold degeneracy into the two double degenerate levels (Kramer's pairs) separated by a gap\cite{Varela2016}. In rings\cite{BercheChatelainMedina}, the degeneracy remains in the fourfold for $\lambda\ne 0$.

\begin{table}[]
\caption{\color{black}Energies (in units of $|t|$) for the first three levels and their degeneracies according to values of the angular momentum label $n$, the electron propagation direction $\nu$, and the spin label $s$ in the limit of
zero spin orbit and zero radiation. Also shown is how degeneracies occur when the SO interaction is turned on, leaving only time reversal symmetry for Kramers pairs.}
\begin{tabular}{ccccl}
\hline
\multicolumn{1}{c}{$E_{n,s}^{\nu,\lambda=0}/|t|$} & \multicolumn{1}{c}{$n$} & \multicolumn{1}{c}{$\nu$} & \multicolumn{1}{c}{$s$} & $\lambda\neq 0$ \\ \hline
0  & -1/2 & + & + & Deg\\
0  & 1/2 & - & - &\\
\cline{2-5} 
0  &  1/2 & + & - & \\
0  & -1/2  & - & + &\\ 
\hline
2  & 1/2  & + & + & Deg\\
2  & -1/2  & - & - &\\ 
\cline{2-5} 
2  & 3/2  & + & - & Deg\\
2  & -3/2  & - & + &\\ 
\hline
4  & 3/2  & + & + & Deg\\
4  & -3/2  & - & - &\\ 
\cline{2-5} 
4  & 5/2  & + & - & Deg\\
4  & -5/2  & - & + &\\ 
\hline
\end{tabular}
\label{table:1}
\end{table}
 
{\color{black}\noindent The corresponding eigenstates are {\color{black} (see appendix)}
 \begin{equation}
 \ket{\psi_s}=\frac{e^{in\varphi}}{\sqrt{2\epsilon}}\left(
\begin{array}{c}
e^{-i\varphi/2}\sqrt{\epsilon+stJ_0(\xi)}       \\
-is e^{i\varphi/2}\sqrt{\epsilon-s tJ_0(\xi)}      
\end{array}     
\right),
 \end{equation}
where $\epsilon=\sqrt{(tJ_0(\xi))^2+4\lambda^2}.$ The system is initially prepared in a general superposition state of the Hamiltonian eigenstates

{\color{black} 
\begin{equation}
\ket{\Psi}=\alpha_+\ket{\psi_+}+\alpha_-\ket{\psi_-},    
\end{equation} 
where the expansion coefficients are subject to the normalization condition $|\alpha_+|^2+|\alpha_-|^2=1$.
Upon substitution of the eigenstate defined previously
\begin{equation}
\ket{\psi_s}=\left(\begin{array}{c}A_s \\B_s
\end{array}\right), 
\end{equation}
we get for $\langle s_z\rangle=\bra{\Psi}s_z\ket{\Psi}$
\begin{equation}
\langle s_z\rangle=|\alpha_+|^2s_{++}+|\alpha_-|^2s_{--}+\alpha_{+}^{*}\alpha_-s_{+-}+\alpha_-^{*}\alpha_+s_{-+},
\end{equation}
with $s_{ab}=\bra{\psi_a}s_z\ket{\psi_b}$, which has the explicit form
\begin{equation}
s_{ab}=\begin{pmatrix}
     A_{a}^{*}&B_{a}^{*}  
\end{pmatrix}\begin{pmatrix}
     A_{b}\\-B_{b}
\end{pmatrix}=A_{a}^{*}A_{b}-B_{a}^{*}B_{b},   
\end{equation}
giving, for instance
\begin{equation}
s_{++}=\frac{(\epsilon+h)-(\epsilon-h)}{2\epsilon}=\frac{h}{\epsilon}.
\end{equation}
where $h=tJ_0(\xi)$ (see appendix). Doing the explicit calculation, one gets
\begin{equation}
\langle s_z\rangle=(|\alpha_+|^2-|\alpha_-|^2)\frac{h}{\epsilon}+\frac{4\lambda}{\epsilon}\Re\alpha_{+}^{*}\alpha_-.
\end{equation}
In some experimental setups in the quantum optics, the states are prepared in such a way that $\alpha_+=\sqrt{p}$, with $p$ being the probability of detection of a given state. Thus, using this insight for a possible experimental detection, one would have to assess:
\begin{equation}
\langle s_z\rangle=(2p-1)\frac{h}{\epsilon}+\sqrt{p(1-p)}\frac{4\lambda}{\epsilon}.
\end{equation}
One can see that the change of sign of $h=tJ_0(\xi)$ can flip the spin expectation value. In the next section we will see how this effect is associated with the split the degenerate Kramers pairs.
} 
We can check whether the pseudospin degree of freedom associated to each strand of the helix is relevant. To see this, we calculate now the inclusion of such hopping term as it is given in the second term of Eq. ($104$) in ref.[\onlinecite{Varela2016}].  Upon inclusion of the term we  
we get the additional pseudospin term $\mathcal{V}_{sp}=t_o J_0(\xi)\sigma_x\mathbbm{1}_s$.
 Since the additional term is diagonal in spin and also is it independent of the orbital angular momentum, we show it just contributes an energy shift. Upon diagonalization, we get the new energies to be given as
  \begin{equation}
  \tilde{E}_{n,s}^{\nu,\sigma}=-{2\nu t}nJ_0(\xi)+s\nu\sqrt{(tJ_0(\xi))^2+4\lambda^2}+\sigma t_oJ_0(\xi),     
\label{spectrum2}
 \end{equation}
with $\sigma=\pm1$. Thus, the minimal Hamiltonian given in equation (\ref{EnergyFull}) does indeed capture the relevant long wavelength physical scenario and we will not in the following, include the pseudospin (strand) degree of freedom. }

\begin{figure}[h]
  	  \centering
  	  \includegraphics[width=8.7cm]{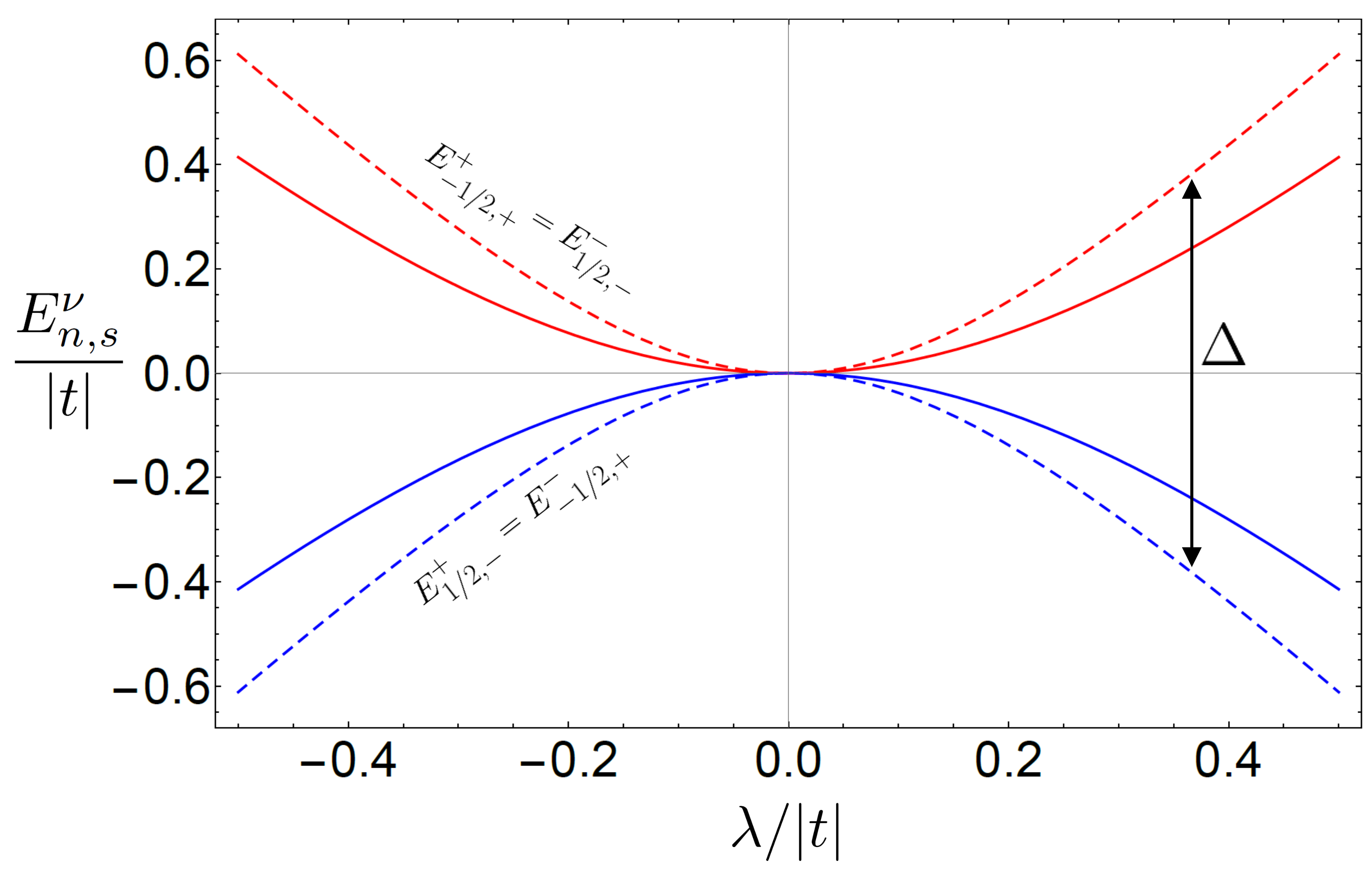}
  	  \caption{Energy of electrons as a function of SO interaction magnitude, $\lambda$ (in units of $t$), with effect of radiation ($\xi=1.5$). The radiation effects thus enhance the gap between helicity states as if an effective additional SO coupling.}
    \label{Figure5}
\end{figure}

\section{Radiation induced helicity splitting}

\begin{figure}[h]
  	  \centering
  	  \includegraphics[width=8.7cm]{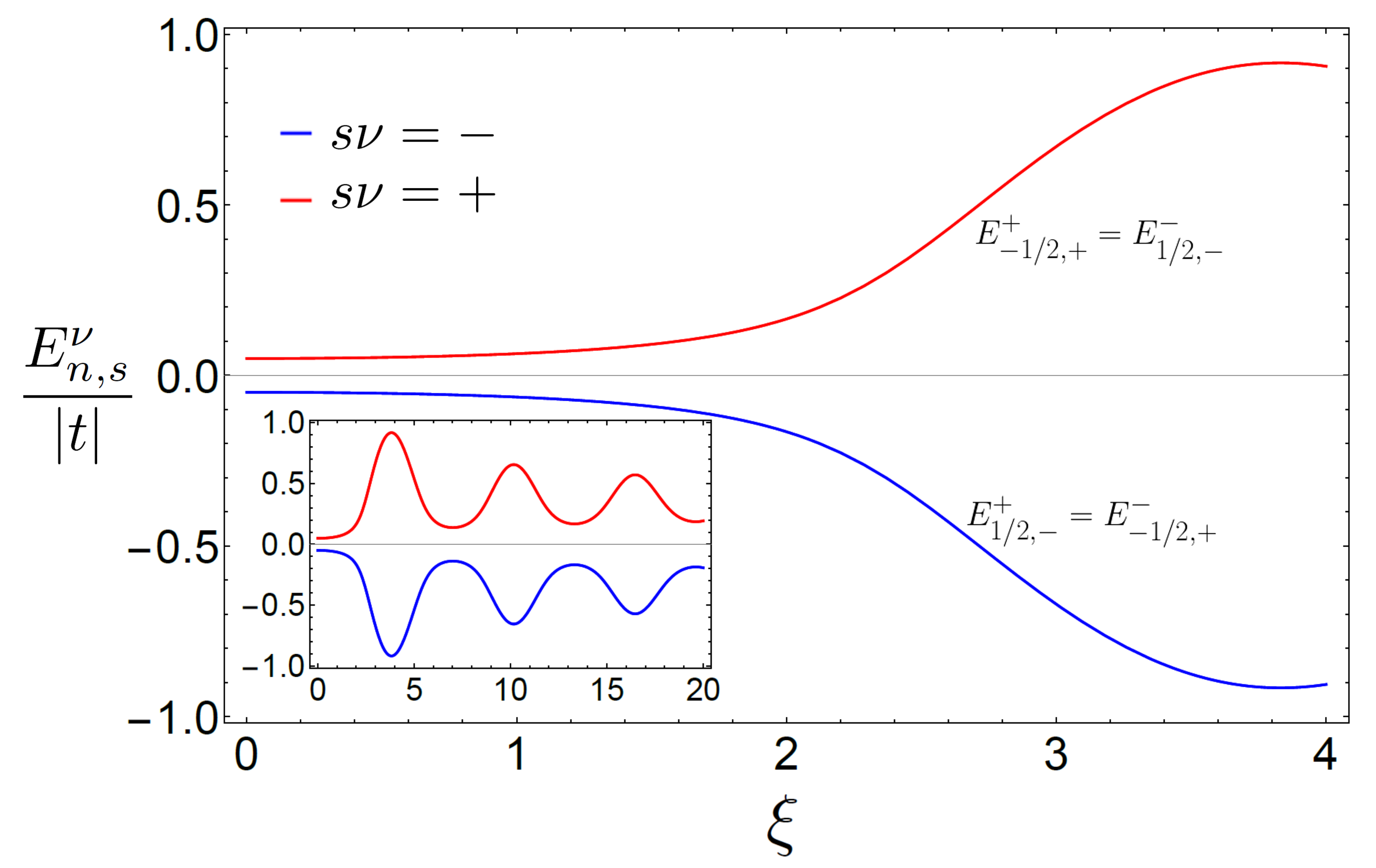}
  	  \caption{\color{black}Energy spectrum in the vicinity of the Fermi level, as a function of $\xi$, for the two helicities $s\nu$. With the SO at $1$ meV coupling present, both helicity states remain almost degenerate until a threshold value for $\xi$ is reached where the Bessel function changes sign inverting the sign of the velocity (see Fig.\ref{EnergyVSk_2}) and enhancing the apparent SO coupling and the helicity splitting. }
  	 
  	  \label{Figure6}
\end{figure}

As seen in Fig.\ref{Figure4} the SO interaction splits the degeneracy between the two helicities in the model. Once the $\lambda\ne0$, radiation can enhance this gap significantly. Figure \ref{Figure5} shows the spectrum in the vicinity of the fermi level as a function of the spin-orbit coupling and radiation intensity. Note how the SO coupling separates the two helicity values introducing a gap between two time reversal symmetric bands of different helicity $\nu s$, given by
\begin{equation}
    \Delta=2tJ_{0}(\xi) + 2\sqrt{(tJ_{0}(\xi))^2 + 4\lambda^{2}}.
    \label{gapEquation}
\end{equation}
Also, once the SO coupling gap is present, it can be enhanced by the radiation intensity. The existence of this gap, beyond the SO limited value, is important in the tunneling problem, very relevant to transport in molecules\cite{VarelaIskra,Aharony,VarelaResponse}. An analogous effect can be induced by uniaxial deformations although they show up by enhancing the SO directly which is more limited in range. This modulation, by the circularly polarized radiation, is capable of producing an helicity splitting in the range of physical parameters that can be very substantial. The energy (Eq.\ref{EnergyFull}) as a function of the radiation $\xi$, is shown in Fig.\ref{Figure6}, with a physical values of $\lambda$. All states are almost degenerate when $\xi=0$, and this quasi-degeneracy persists up to a value of $\xi$ for which we have the first node of the Bessel function $J_0(\xi)$. In the region between the first and second nodes, we observe that radiation induces a gap enhacement between states of different helicity, having an effect of generating an additional effective SO interaction on these states. This enhancement is more than ten-fold in the Figure. The gap between helicity states can also be derived from Eq.\ref{gapEquation}. The behavior of the states for higher radiation values is shown in the inset, where it is observed that the gapped regions oscillate with the field intensity/frequency. However, time reversal symmetry is always satisfied, i.e. $E^{\nu}_{n,s}=E^{\nu}_{-n,-s}$ (simultaneous change of $\lambda$ and s). The effect of the SO interaction on the energy together with radiation effects is shown in Fig.\ref{Figure6}.


\section{Summary and conclusions}

We studied a model of double stranded DNA, where mobile electrons live on the $\pi$ orbitals of the bases and the spin-orbit coupling is first order (nearest neighbors) due to geometry\cite{Varela2016}. 
The Hamiltonian model contains both a kinetic term for intra and inter-strand terms and spin-orbit couplings that are only first order for intra-strand couplings. To this model we a apply circularly polarized radiation along the axis of the molecule. By way of the Floquet formalism we obtain an effective Hamiltonian valid in the limit of high frequency $\hbar\Omega>t_i$ the inter strand hopping term of the unperturbed model.

Focusing on the half filled model we find a radiation dependent kinetic term plus a spin couple term that depends on the radiation parameter $\xi=eER/\hbar\Omega$ where $E$ is the electric field amplitude, $\Omega$ is the radiation frequency and $R$ is the inter-base distance. We have carefully derived the eigenvalues and eigenfunctions of the radiation-spin-orbit coupled system around half filling and find somewhat similar spectral behavior as for mesoscopic rings in the absence of radiation except for the quantization of the wave-vector in the latter case and the degeneracy of the ground states. In contrast to the one dimensional Rashba model\cite{Winkler,VarelaIskra}, where there is no associate angular momentum, we find that the spin-orbit coupling breaks the degeneracy of different helicity states. 

When radiation is turned on the kinetic energy gets modulated with $\xi$ and can change sign within reasonable physical values of the parameters. The radiation also produces an effective SO coupling once the intrinsic coupling is finite that can enhance the helicity splitting gap even tenfold. Such helicity splitting in the spectrum is relevant to tunneling and polaron transport problems and the Chiral Induced Spin Selectivity effect.

\acknowledgements{ The authors thank V. Mujica for useful discussions. This work is part of the project: Mechanical Spectroscopy of chiral molecules: Force and radiation interactions at Yachay Tech.}

\section{Data Availability Statement}

The data that support the findings of this study are available from the corresponding author upon reasonable request.

\appendix\label{appendix}

\section{On Floquet Fourier Approach}\label{app:ffa}
To complement the former discussion, let us explicitly show more details about the calculation of the effective Floquet Hamiltonian. Hence, we consider a generic time dependent periodic Hamiltonian $H(t+T)=H(t)$. We write it as free Hamiltonian plus a time dependent interaction
\begin{equation}\label{1}
H(t)=H_0+V(t) 
\end{equation}
and we assume that we can solve the dynamics of the free part $H_0$. Its eigenbasis is spanned by the spinors $|\varphi_{\alpha}\rangle$ 
where $\alpha$ describes a set of quantum numbers. We now use the eigenstates $|\varphi_{\alpha}\rangle$ as expansion basis for the eigenstates 
of the full Hamiltonian in Eq.\ref{1}. 

In order to analyze the evolution equation 
\begin{equation}
i\hbar\partial_t|\varphi(t)\rangle=H(t)|\varphi(t)\rangle.\label{2}
\end{equation}
we take advantage of the periodicity of the Hamiltonian so we can resort to Floquet's theorem.
For this purpose we define an auxiliary hermitian Hamiltonian 
\begin{equation}
\mathcal{H}(t)=H(t)-i\hbar\partial_t,
\end{equation}
along with the so called Floquet states 
\begin{equation}
|\Psi_{\alpha}(t)\rangle=\exp(i\varepsilon_{\alpha} t)|\varphi(t)\rangle
\end{equation}
such that
\begin{equation}\label{evolution}
\mathcal{H}(t)|\Psi_{\alpha}(t)\rangle=\varepsilon_{\alpha}|\Psi_{\alpha}(t)\rangle 
\end{equation}
which are periodic functions of time, 
$|\Psi_{\alpha}(t+T)\rangle=|\Psi_{\alpha}(t)\rangle$. In addition, the eigenvalues $\varepsilon_{\alpha}$ form the quasi-energy spectrum, and are the analogous of the quasi-momenta for Bloch electrons in a spatially periodic structure. Since the states 
\begin{equation}
|\Psi_{\alpha n}(t)|\rangle=\exp(in\Omega t)|\Psi_{\alpha}(t)\rangle 
\end{equation}
are also eigenstates of the Hamiltonian $\mathcal{H}(t)$ but with corresponding eigenvalues 
$\varepsilon_{\alpha}\rightarrow\varepsilon_{\alpha}+n\Omega$, we 
can work in the {\it first Brillouin zone} $-\Omega/2\leq\varepsilon_{\alpha}\leq\Omega/2$. 

Using the periodic temporal basis  $\xi_n(t)=\exp(in\Omega t)$, which satisfies
\begin{displaymath}
\frac{1}{T}\int^{T}_0\xi^{*}_n(t)\xi_m(t)=\delta_{nm},
\end{displaymath}
we write the Fourier mode expansion
\begin{equation}
\varphi_{\alpha}(t)=\exp({-i\varepsilon_{\alpha }t})\sum^{n=\infty}_{n=-\infty}|C^{(n)}_{\alpha}\rangle\xi_n(t),
\end{equation}
Now we use the expansion $|C^{(n)}_{\alpha}\rangle=\sum_{\beta}\Lambda^{(n)}_{\alpha\beta}|\varphi_{\beta}\rangle$ 
such that the Eq.(\ref{evolution}) becomes
\begin{eqnarray}
H(t)\sum^{n=\infty}_{n=-\infty}\sum_{\beta}\Lambda^{(n)}_{\alpha\beta}|\varphi_{\beta}\rangle\xi_n(t)+\\\nonumber
\sum^{n=\infty}_{n=-\infty}\sum_{\beta}\Lambda^{(n)}_{\alpha\beta}|\varphi_{\beta}\rangle\xi_n(t)(\varepsilon_{\alpha}-n\Omega)=0\label{eigenvalue}.
\end{eqnarray}
Multiplication by $\langle\varphi_{\gamma}|\xi^{*}_{m}(t)$, then average over one temporal period, leads to
\begin{equation}
\sum^{n=\infty}_{n=-\infty}\sum_{\beta}[\langle\alpha|H^{(m-n)}|\beta\rangle-(\varepsilon_{\alpha}-m\Omega)\delta_{nm}\delta_{\alpha\beta}]\Lambda^{(n)}_{\alpha\beta}=0.
\end{equation}
and we have used the simplifying notation $|\alpha\rangle\equiv|\varphi_{\alpha}\rangle$, and $H^{(m-n)}=1/T\int^{T}_0\xi^{*}_m(t)H(\mathbf{k},t)\xi_n(t)$. 

Then, the quasienergies $\varepsilon_{\alpha}$ are eigenvalues of the secular equation.
\begin{equation}
det|H_F-\varepsilon_{\alpha}|=0\label{secular}
\end{equation}
where $\langle\langle\alpha m|H_F|n\beta\rangle\rangle=H^{(m-n)}_{\alpha\beta}+m\Omega\delta_{nm}\delta_{\alpha\beta}$. 
Here we have used 
$|\alpha m\rangle\rangle\equiv|\alpha\rangle\otimes\xi(t)$, which  represents the direct product of orbital and periodic eigenfunctions.

\section{2D Hamiltonian derivation}\label{AppendixB}
The full tight-binding Hamiltonian is
\begin{equation}\label{FullHamitonian}
\mathcal{H} = t_i\sum_{\imath\jmath}^{intra}c_{\imath}^{\dagger}c_{\jmath} + i\lambda\sum_{\imath\jmath}^{intra}c_{\imath}^{\dagger}\nu_{\imath\jmath}s_\varphi c_{\jmath} .
\end{equation}
 
In the Bloch space, the Hamiltonian is: 
\begin{equation}
H=\left(\epsilon_{2p}^{\pi} + 2t_i f(k)\right)\mathbf{1}_{\sigma}\mathbf{1}_s - 2\lambda g(k)\mathbf{1}_{\sigma}\mathbf{s_\varphi}.
\end{equation}
with $f(\mathbf{k})=\cos(\mathbf{k}\cdot \mathbf{R})$ and $g(\mathbf{k})=\sin(\mathbf{k}\cdot\mathbf{R})$, being $R$ the lattice parameter\cite{Varela2016}.  
\\

The position vector for sites in the chain is given by
\begin{equation}
\mathbf{R}=R_x \hat{x} + R_y \hat{y} + R_z \hat{z};
\end{equation}
and we can see of the figure \ref{Figure1} that the following relation is true: 
\begin{equation}
\tan{\eta}=\frac{R_z}{R_y}=\frac{k_z}{k_y},
\end{equation}
and the general wave vector in the reciprocal space is $\mathbf{k}=k_x\hat{x}+k_y\hat{y}+k_z\hat{z}$. 
\\

Also, for the kinetic energy the eigenvalues  are
\begin{equation}\label{energy}
E=\frac{\epsilon_{2p}^{\pi}+2t^{in}f(k)\mp t^{out}}{1+f(k)s_o\mp \gamma_{o}}. 
\end{equation}
We consider the approximation where $s_o=\gamma_o=0$.  

In the half filling case, using equation (\ref{energy}), and $\mathbf{k}\cdot\mathbf{R}=\nu\pi/2$ with $\nu=\pm 1$, the $\mathbf{K}_{\nu}$ points are
\begin{equation}
\mathbf{K_{\nu}}=\nu\left(0,\frac{\pi}{2R\sec{\eta}},\frac{\pi}{2R\csc{\eta}}\right). 
\end{equation}
Considering a small perturbation $q$ around $\mathbf{K_{\nu}}$ in form $\mathbf{k}=\mathbf{K_{\nu} +\mathbf{q}}$, the functions are
\begin{equation}
f(\mathbf{k}) \approx  -\nu(q_y + q_z\tan{\eta})R\cos{\eta}+...
\end{equation}
and 
\begin{equation}
g(\mathbf{k}) \approx \nu + ...
\end{equation}
Substituting in (\ref{FullHamitonian}), the Hamiltonian is
\begin{equation}
H=[\epsilon_{2p}^{\pi} - 2\nu t^{in}(q_y + q_z\tan{\eta})R\cos\eta]\mathbf{1}_{s} - 2\nu\lambda_{SO}^{in}\mathbf{s_\varphi}, 
\end{equation}
or, in term only of $q_y$
\begin{equation}
H=(\epsilon_{2p}^{\pi} - 2\nu t^{in}q_yR\sec{\eta})\mathbf{1_{s}} -2\nu\lambda_{SO}^{in}\mathbf{s_\varphi}.
\end{equation}


\section{Eigenfunctions}

We start by writing the eigenvalue equation for the spin-orbit dependent contribution to the Hamiltonian in eq. (11), i.e. $H=hs_z-2\lambda s_\varphi$, where we have put $h=tJ_0(\xi)$ for simplicity of the forthcoming calculations. Then,$H\ket{\psi_s}=\epsilon_s\ket{\psi}$ leads to $H^2 \ket{\psi_s}=\epsilon_s^2\ket{\psi_s}\rightarrow h^2+4\lambda^2=\epsilon_s^2$, which in turn implies that $\epsilon_s=s\sqrt{h^2+4\lambda^2}.$ The calculation of the eigenstates goes like this:
We write the eigenvalue equation as
\begin{equation}\label{ev1}
\left(
\begin{array}{cc}
     h  &2i\lambda e^{-i\varphi}\\
2i\lambda e^{i\varphi}     &-h 
\end{array}
\right)\left(\begin{array}{c}A_s \\B_s
\end{array}\right)=\epsilon_s\left(\begin{array}{c}A_s \\B_s
\end{array}\right),
\end{equation}
with $\epsilon_s=s\sqrt{h^2+4\lambda^2}=s\epsilon$ and the normalization condition $|A_s|^2 +|B_s|^2=1$. From the eigenvalue equation, we get
\begin{equation}
hA_s+2i\lambda B_s=\epsilon_sA_s    
\end{equation}
which in turn gives the relation:
\begin{equation}
\frac{B_s}{A_s}=\frac{\epsilon_s-h}{2i\lambda}.    
\end{equation}
Upon substitution on the normalization condition, we get
\begin{equation}
|A_s|^2\left(1+\frac{(\epsilon_s-h)^2}{4\lambda^2}\right)=1    
\end{equation}
which is equivalent to
\begin{equation}
|A_s|^2(4\lambda^2+\epsilon_s^2-2h\epsilon_s+h^2)=4\lambda^2    
\end{equation}
or 
\begin{equation}
|A_s|^2(2\epsilon^2-2sh\epsilon)=4\lambda^2    
\end{equation}
which in turn leads to
\begin{equation}
|A_s|=\sqrt{\frac{\epsilon+sh}{2\epsilon}}.    
\end{equation}
In order to choose the phase, we take into account that $\psi_s$ ought to be also eigenstate of the total angular momentum. Therefore, we set
\begin{equation}
A_s=\frac{e^{in\varphi}}{\sqrt{2\epsilon}}e^{-i\phi/2}\sqrt{\epsilon+sh}.    
\end{equation}
Using this result, we get for the other component of the eigenstate
\begin{equation}
B_s=-is\frac{e^{in\varphi}}{\sqrt{2\epsilon}}e^{i\phi/2}\sqrt{\epsilon-sh}.    
\end{equation}
Thus, the eigenstate reads as
 \begin{equation}
 \ket{\psi_s}=\frac{e^{in\varphi}}{\sqrt{2\epsilon}}\left(
\begin{array}{c}
e^{-i\varphi/2}\sqrt{\epsilon+stJ_0(\xi)}       \\
-is e^{i\varphi/2}\sqrt{\epsilon-s tJ_0(\xi)}      
\end{array}     
\right),
 \end{equation}




%



\end{document}